\documentclass{emulateapj}

\usepackage{natbib}

\slugcomment{To Appear in the Astronomical Journal}

\shorttitle{SN 2013df and Its Progenitor}
\shortauthors{Van Dyk et al.}

\begin{document}

\title{The Type IIb Supernova 2013df and Its Cool Supergiant Progenitor}

\author{Schuyler D.~Van Dyk\altaffilmark{1},
  WeiKang Zheng\altaffilmark{2},
  Ori D.~Fox\altaffilmark{2},
  S.~Bradley Cenko\altaffilmark{3},
  Kelsey I.~Clubb\altaffilmark{2},
  Alexei V.~Filippenko\altaffilmark{2},
  Ryan J.~Foley\altaffilmark{4,5},
  Adam A.~Miller\altaffilmark{6,7}, 
 Nathan Smith\altaffilmark{8},
 Patrick L.~Kelly\altaffilmark{2},
 William H.~Lee\altaffilmark{9},
 Sagi Ben-Ami\altaffilmark{10},
 and Avishay Gal-Yam\altaffilmark{10}
}

\altaffiltext{1}{Spitzer Science Center/Caltech, Mail Code 220-6,
  Pasadena, CA 91125; email: vandyk@ipac.caltech.edu.}
\altaffiltext{2}{Department of Astronomy, University of California,
  Berkeley, CA 94720-3411.}
\altaffiltext{3}{Astrophysics Science Division, NASA Goddard Space Flight Center, Mail Code 661, 
Greenbelt, MD 20771.}
\altaffiltext{4}{Harvard-Smithsonian Center for Astrophysics, 60 Garden Street, Cambridge, MA 02138.}
\altaffiltext{5}{Department of Astronomy, University of Illinois, Urbana-Champaign, IL 61801.}
\altaffiltext{6}{Jet Propulsion Laboratory, MS 169-506, Pasadena, CA 91109.}
\altaffiltext{7}{Hubble Fellow.}
\altaffiltext{8}{Steward Observatory, University of Arizona,
  Tucson, AZ 85720.}
\altaffiltext{9}{Instituto de Astronom\'ia, Universidad Nacional Aut\'onoma de M\'exico, 
Apdo.~Postal 70-264, Cd.~Universitaria, M\'exico DF 04510, M\'exico}
\altaffiltext{10}{Benoziyo Center for Astrophysics, The Weizmann Institute of Science, Rehovot 76100, 
Israel.}

 \begin{abstract}
We have obtained early-time photometry and spectroscopy of Supernova (SN) 2013df in NGC 4414. 
The SN is clearly of Type IIb, with notable similarities to SN 1993J. 
From its luminosity at secondary maximum light, it appears that 
less $^{56}$Ni ($\lesssim 0.06\ M_{\odot}$)
was synthesized in the SN 2013df explosion than was the case for the SNe~IIb 1993J, 2008ax, and 
2011dh. Based on a comparison of the light curves, the SN 2013df progenitor must have been more 
extended in radius prior to explosion than the progenitor of SN 1993J.
The total extinction for SN 2013df is estimated to be $A_V=0.30$ mag. The metallicity at the SN
location is likely to be solar.
We have conducted {\sl Hubble Space Telescope\/} ({\sl HST}) Target of Opportunity
observations  of the SN
with the Wide Field Camera 3, and from a precise comparison of these new observations to archival 
{\sl HST\/} observations of the host galaxy obtained 14 years prior to explosion, 
we have identified the progenitor of SN 2013df to be a yellow
supergiant, somewhat hotter than a red supergiant progenitor for a normal Type II-Plateau SN. 
From its observed spectral energy distribution, assuming that the light is dominated by one star,
the progenitor had effective temperature $T_{\rm eff} = 4250 \pm 100$ K and a bolometric
luminosity $L_{\rm bol}=10^{4.94 \pm 0.06}\ L_{\odot}$. This leads to an effective radius 
$R_{\rm eff} = 545 \pm 65\ R_{\odot}$.
The star likely had an initial mass in the range of 13--17 $M_{\odot}$; however, if it was 
a member of an interacting binary system, detailed modeling of the system is required to estimate 
this mass more accurately.
The progenitor star of SN 2013df appears to have been relatively similar to the progenitor of SN 1993J.
 \end{abstract}

\keywords{ galaxies: individual (NGC 4414) --- stars: evolution
--- supernovae: general --- supernovae: individual (SN 2013df)}

\section{Introduction}\label{intro}

\bibpunct[;]{(}{)}{;}{a}{}{;} 

A core-collapse supernova (SN) is the final stage of evolution for stars with 
initial mass $M_{\rm ini} \gtrsim 8\ M_{\odot}$ \citep[e.g.,][]{woosley02}. 
The compact remnant of this explosion is thought to 
be either a neutron star or a black hole.
Stars which reach the terminus of their evolution with most of their hydrogen-rich envelope intact --- single red supergiants (RSGs) ---
produce Type II (specifically Type II-Plateau) supernovae (SNe).
As the envelope is stripped away, either through vigorous mass loss 
or via mass transfer in a binary system, the results are 
the hydrogen-free, yet helium-rich, Type Ib SNe and the hydrogen-free and helium-poor 
(or helium-free) Type Ic SNe; see \citet{filippenko97} for a review
of SN classification.
Intermediate to the SNe~II and SNe~Ib are the Type IIb, which retain a 
low-mass ($\lesssim 1\ M_{\odot}$) hydrogen envelope prior to explosion.
SNe~IIb are
intrinsically rare \citep[$\sim 10$--11\% of all core-collapse SNe;][]{smith11,li11}.
The first identified case was SN 1987K \citep{filippenko88}; well-studied 
examples are the nearby 
SN~1993J in Messier~81 \citep[e.g.,][]{richmond96,matheson00} and SN~2011dh in Messier~51
\citep[e.g.,][]{arcavi11,ergon13}. 
The progenitor channel for SN 1993J which has garnered most favor 
over the years has been a massive interacting binary system 
\citep[e.g.,][]{pod93,woosley94,maund04,stancliffe09}. A similar model has been introduced for
SN 2011dh \citep{benvenuto13}.
The progenitor is the mass donor via Roche-lobe overflow.

Spectroscopically, SNe~IIb exhibit at early times emission and absorption features 
due to H, resembling SNe~II, which subsequently give way to the emergence of the He~{\sc i} 
absorption lines typical of SNe~Ib \citep[e.g.,][]{filippenko93,chornock11}, 
with broad H emission reappearing in the nebular phase
\citep[e.g.,][]{filippenko94,matheson00,taubenberger11,shivvers13}.
Although rarely seen owing to its very short duration early in the evolution of a SN, 
a rapid decline after an initial peak has been observed among SNe~IIb 
\citep[e.g.,][]{richmond94,richmond96,roming09,arcavi11}. This is interpreted as 
adiabatic cooling after the SN shock has broken 
out through the star's surface; the duration of this cooling is governed primarily by the radius of
the progenitor \citep{chevalier08,nakar10, rabinak11,bersten12}. The SN light curves reach a 
secondary maximum,
as a result of thermalization of the $\gamma$-rays and positrons emitted during the radioactive decay
of $^{56}$Ni and $^{56}$Co, followed by a smooth decline. The light-curve shapes resemble those of
SNe~Ib \citep{arcavi12}.

As the number of all core-collapse SNe having directly identified
progenitor stars is very small ($\sim 20$ at the time of this writing), 
we have been extraordinarily fortunate that, up to this point, the progenitors of three SNe~IIb have 
been identified, including SN~1993J \citep{aldering94,vandyk02,maund04,maund09}, 
SN~2008ax in NGC~4490 \citep{crockett08}, and 
SN~2011dh in Messier~51 \citep{maund11,vandyk11}.
Each of these stars exhibits distinctly different properties, although all show indications of 
envelope stripping prior to explosion: the SN 1993J progenitor has been characterized
as a K-type supergiant, with initial mass $\sim 13$--$22\ M_{\odot}$
\citep{vandyk02,maund04}; for SN 2008ax it was difficult to fit a single supergiant to the 
observed colors \citep{li08,crockett08}, and the possible initial mass range for the progenitor
is large, $\sim 10$--$28\ M_{\odot}$; and for SN 2011dh we now know that the $\sim 6000$~K 
yellow supergiant identified by \citet{maund11} and \citet{vandyk11} was indeed the star that 
exploded \citep{ergon13,vandyk13b}, with initial mass $\sim 12$--$15\ M_{\odot}$
\citep{maund11,murphy11,bersten12}.
In addition, \citet{ryder06} may have detected at very late times the massive companion to
the SN IIb~2001ig in NGC 7424 \citep{silverman09}.
Furthermore, \citet{chevalier10} have divided SN~IIb progenitors into those that are
compact (radius $\sim 10^{11}$ cm; e.g., SN 2008ax) and 
those that are extended ($\sim 10^{13}$ cm; SN 1993J), although
SN 2011dh appears to be an intermediate case \citep{horesh13}.
Each new example provides us with an increased understanding of this SN subtype and of the
massive stars which give rise to these explosions.

\begin{figure}
\figurenum{1}
\plotone{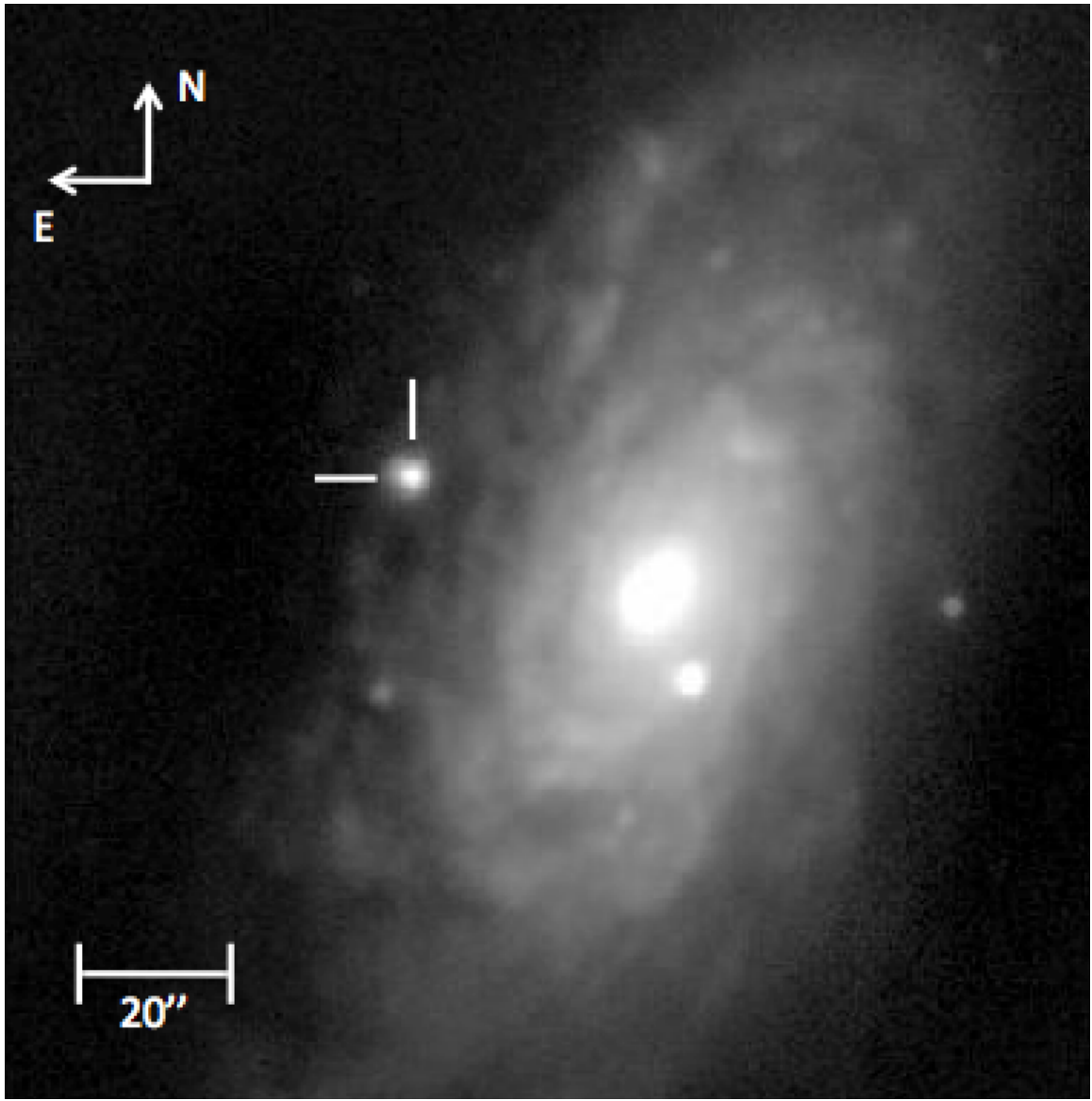}
\caption{A greyscale composite of $rizJH$ images obtained with RATIR on 2013 June 24, showing 
SN 2013df in NGC 4414. 
The position of the SN is indicated by the tick marks. North is up, east is to the left.\label{figsn}}
\end{figure}

In this paper we consider SN~2013df in NGC~4414, shown in Figure~\ref{figsn}.
It was discovered by \citet{ciab13} on June 7.87 and 8.83 
(UT dates are used throughout).
\citet{cenko13} provided
spectroscopic confirmation on June 10.8 that it is a Type II SN. The resemblance to SN~1993J at early times
suggested to \citeauthor{cenko13}~that SN~2013df would evolve to be a SN~IIb.
Here we present early-time photometric and spectroscopic observations of SN~2013df, which 
demonstrate that it is indeed a SN IIb. In a telegram,
we presented a preliminary identification of three progenitor candidates for the SN \citep{vandyk13a}.
We provide here a much better identification, through high-resolution imaging of the SN, and show that 
none of the three sources turned out actually to be at the SN position.
We will characterize the nature of the probable progenitor.
The host galaxy, NGC~4414, is a nearby, isolated, flocculent spiral galaxy, with an inclination of 
$55\arcdeg$ \citep{vallejo02}. \citet{thornley97} measured a global star formation rate of 
1.3 $M_{\odot}$ yr$^{-1}$, comparable to that of other Sc galaxies.
SN~2013df is at a nuclear offset of 32\arcsec\ E, 14\arcsec\ N, along an
outer spiral arm.
The galaxy was also host to the likely Type Ia SN~1974G \citep[e.g.,][]{schaefer98}.
We adopt the Cepheid-based distance modulus $\mu_0 = 31.10 \pm 0.05$ mag 
(distance $16.6 \pm 0.4$ Mpc) established by
\citet{freedman01}.

\section{Observations}

\subsection{Early-Time Photometry}

We have observed the SN using the 0.76\,m Katzman Automatic Imaging Telescope 
\citep[KAIT;][]{filippenko01} at Lick Observatory in $BVRI$ between 2013 June 13.7
and July 18.7.
Point-spread function (PSF) photometry was applied using the DAOPHOT \citep{stetson87} 
package from the IDL Astronomy User's Library\footnote{http://idlastro.gsfc.nasa.gov/contents.html.}. 
The instrumental magnitudes and colors of the SN were transformed to standard 
Johnson-Cousins $BVRI$ using two stars in the SN field which are in the Sloan Digital
Sky Survey (SDSS) catalog, and by following this 
prescription\footnote{http://www.sdss.org/dr7/algorithms/sdssUBVRITransform.html\\ \#Lupton2005.}
to convert from the SDSS  to the Johnson-Cousins system.

Data in $rizJH$ were also obtained with the multi-channel Reionization And Transients InfraRed 
camera \citep[RATIR;][]{butler12} mounted on the 1.5-m Harold L.~Johnson telescope at the Mexican 
Observatorio Astron\'omico Nacional on Sierra San Pedro M\'artir in Baja California, M\'exico
\citep{watson12}. Typical observations include a series of 60-s exposures, with dithering 
between exposures. 
RATIR's fixed infrared (IR) filters cover half of their respective detectors, automatically providing off-target IR sky 
exposures while the target is observed in the neighboring filter.  Master IR sky frames were created 
from a median stack of off-target images in each IR filter.  No off-target sky frames were obtained with 
the optical CCDs, but the small galaxy size and sufficient dithering allowed for a sky frame to be 
created from a median stack of all the images in each filter.  Flat-field frames consist of evening sky 
exposures. Given the lack of a cold shutter in RATIR's design, IR ``dark'' frames are not available.  Laboratory 
testing, however, confirms that dark current is negligible in both IR detectors \citep{fox12}. 
The data were reduced, coadded, and analyzed using standard CCD and IR processing techniques 
in IDL and Python, utilizing online astrometry programs {\tt SExtractor} \citep{bertin96} and 
{\tt SWarp}\footnote{SExtractor and SWarp can be accessed from http://www.astromatic.net/software.}.
Calibration was performed using a single comparison star in the SN field that also has reported fluxes in both 
2MASS \citep{skrutskie06} and the SDSS Data Release 9 Catalogue 
\citep{ahn12}. 

We did not yet possess template images of the host galaxy (i.e., prior to the SN discovery or when 
the SN has faded to invisibility) to subtract from the images with the SN present in each 
band. Consequently, the photometry presented here should be considered preliminary; however, the 
SN is far from the main light from the host galaxy, so results including template subtraction prior to
photometry might not substantially differ from what we present here.

\subsection{Early-Time Spectroscopy}

We have obtained a number of spectra of the SN at early times, using the Kast spectrograph
\citep{miller93} on
the Lick Observatory 3-m Shane telescope and the DEep Imaging Multi-Object Spectrograph
\citep[DEIMOS;][]{faber03} on the Keck-II 10-m telescope.
Ultraviolet (UV) spectroscopy has also been obtained using the {\sl Hubble Space Telescope\/} 
({\sl HST}) Space Telescope Imaging Spectrograph (STIS) as part of program GO-13030
(PI: A.~V.~Filippenko). The results of these {\sl HST\/} observations will be presented in a future paper, together with the bulk 
of the ground-based optical spectra. However, here we present and analyze
a representative spectrum obtained on 2013 July 11.2 with DEIMOS using the 600 l mm$^{-1}$ grating.

\subsection{{\sl HST\/} Imaging}

The region of the host galaxy containing the SN site 
was observed with the {\sl HST\/} Wide Field Planetary Camera 2 on 1999 April 29
by program GO-8400 (PI: K.~Noll), as part of the Hubble Heritage Project. The bands used were 
F439W (two individual images with 40~s exposure times and two with 1000~s), 
F555W (four 400-s exposures), F606W (two 60-sec exposures), and 
F814W (two 40-s and four 400-s exposures). The F555W and F814W 
data were combined with images obtained by programs GO-5397 and GO-5972
at an earlier time for the rest 
of the galaxy, and drizzled into mosaics
in each band at the scale $0{\farcs}05$ pix$^{-1}$ by \citet{holwerda05}.

We have also observed the SN on 2013 July 15 with {\sl HST\/}
using the Wide Field Camera~3 (WFC3) UVIS channel in F555W, as part of our Target of 
Opportunity program GO-12888 (PI: S.~Van Dyk).
The observations consisted of 28 5-s exposures; the short exposure time was intended to avoid
saturation in each frame by the bright SN.

\section{Analysis}

\subsection{Light Curves}

\begin{figure}
\figurenum{2}
\plotone{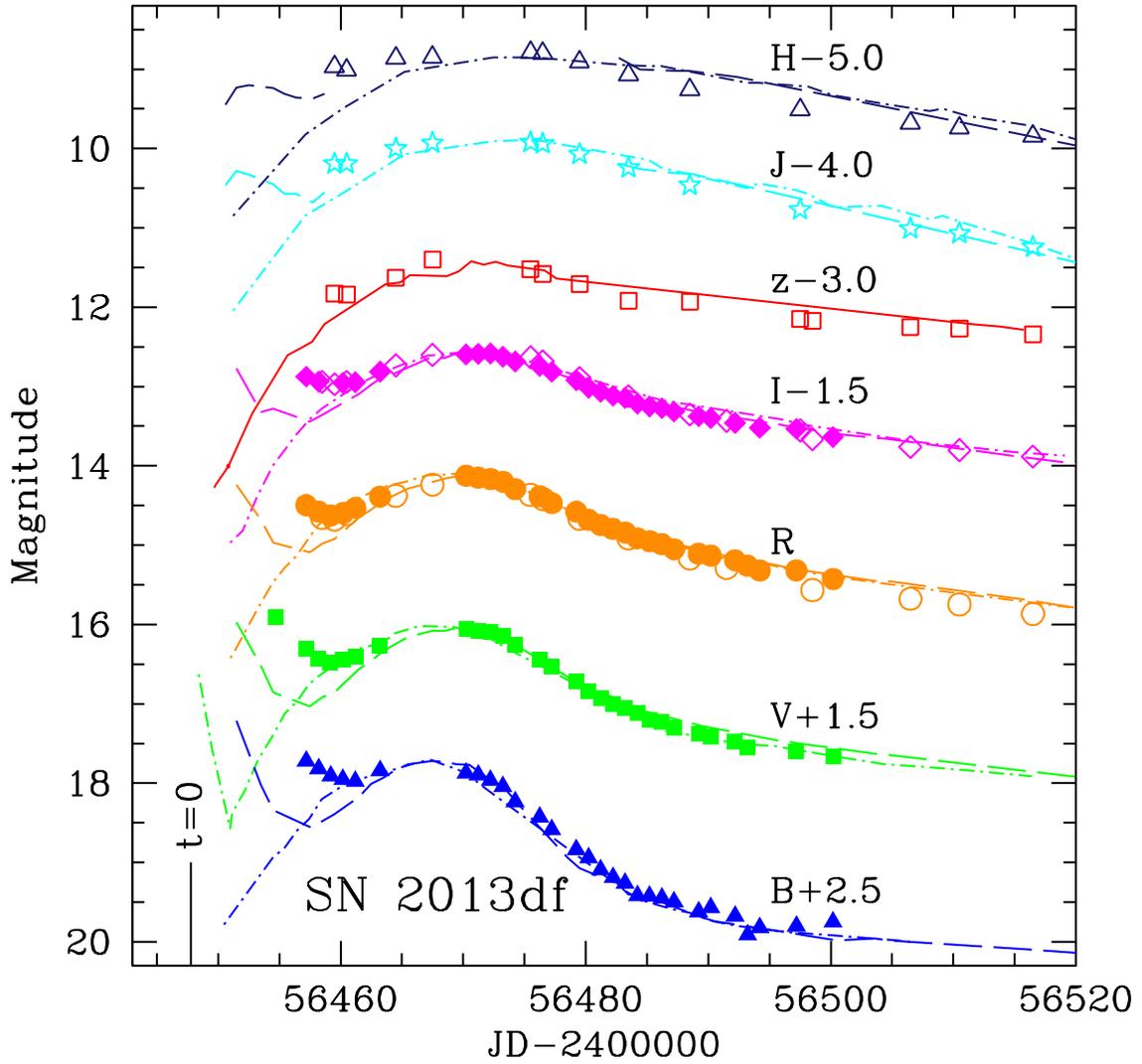}
\caption{$BVRIzJH$ light curves of SN 2013df from KAIT (solid points) and RATIR
(open points; with $r$ and $i$ converted to $R$ and $I$, respectively, following 
http://www.sdss.org/dr7/algorithms/sdssUBVRITransform.html\#Lupton2005). 
The observed curves have been
offset from each other for clarity. Shown for comparison
are the $BVRIJH$
light curves of SN 1993J \citep[][ dashed lines]{richmond94,richmond96,matthews02}
and of SN 2011dh \citep[][ dot-dashed lines]{arcavi11,vandyk13b,ergon13}, and 
the $z$-band light curve of SN 2008ax \citep[][ solid line]{pastorello08}, all 
adjusted in time and relative brightness. The estimated time $t=0$ of the explosion is indicated.
\label{figlc}}
\end{figure}

In Figure~\ref{figlc} we display the early-time KAIT and RATIR light curves in all bands for SN 2013df.
We also include the very early $V$ measurement from June 11.202 by Stan 
Howerton\footnote{http://www.flickr.com/photos/watchingthesky/9035874997.}.
We compare these curves with those at $BVRIJH$ for 
SN 1993J \citep{richmond94,richmond96,matthews02} and for 
SN 2011dh \citep{arcavi11,vandyk13b,ergon13}.
We also compare the $z$-band light curve of SN 2008ax \citep{pastorello08}.
The curves for the comparison SNe were adjusted in time and relative brightness to match the curves of 
SN 2013df, particularly at the secondary maximum in each band.
Clearly, from its overall photometric similarity with the other SNe, SN 2013df appears to be a SN~IIb;
see \citet{arcavi12} for a general description of SN~IIb light-curve shapes.
What is most notable from the comparison is that the post-shock-breakout cooling of 
SN 2013df occurred at a later epoch in all bands, relative to that of SN 1993J. 
The post-breakout decline of SN 2011dh occurred at an even earlier relative epoch 
\citep[e.g.,][]{arcavi11,bersten12}.

\begin{figure}
\figurenum{3}
\plotone{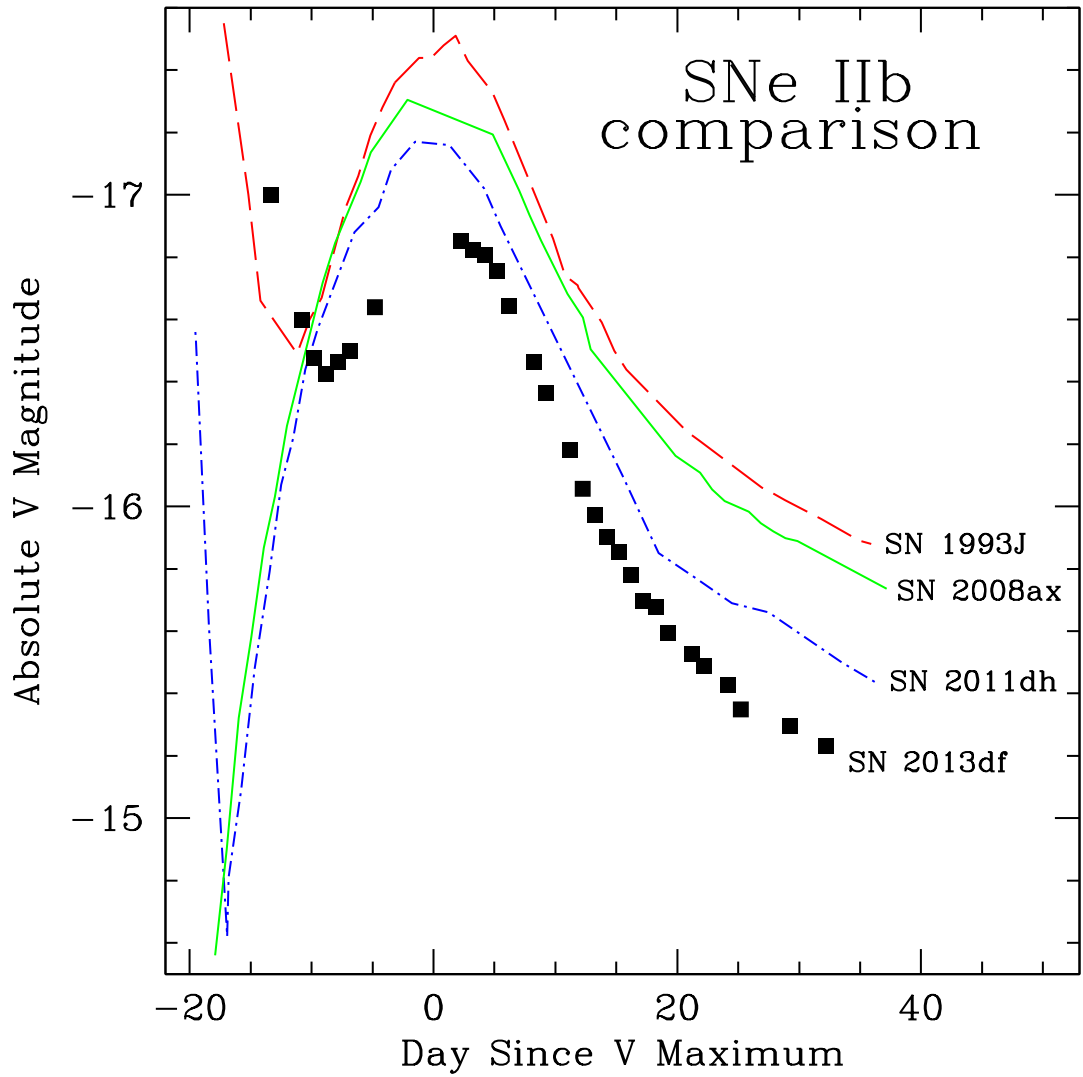}
\caption{Absolute $V$ light curve for SN 2013df (solid points), compared to those at $V$ 
for SNe 1993J (dashed line; colored red in the online version), 
2008ax (solid line; colored green in the online version), and 
2011dh (dot-dashed line; colored blue in the online version) shown in 
Figure~\ref{figlc}.
All of the curves are displayed relative to the day of $V$ maximum; see text.\label{figlcabs}}
\end{figure}

We show in Figure~\ref{figlcabs} the absolute $V$-band light curve of SN 2013df, relative to those of
SNe 1993J, 2008ax, and 2011dh. 
(We do not make this comparison based on bolometric or pseudo-bolometric luminosity, since, as
noted, we consider the present photometry of SN 2013df to be preliminary.)
The light curve of SN 2008ax is from \citet{pastorello08}, rather than from 
\citet{taubenberger11}, since the former has somewhat more complete 
coverage through the secondary maximum. 
The epochs of the secondary maxima in $V$ for SNe 1993J, 2008ax, and 2011dh are from
\citet{richmond96}, \citet{pastorello08}, and \citet{ergon13}, respectively. This epoch for SN 2013df 
was determined by comparing its light curve to that of each of the three comparison SNe.
The light curves have all been adjusted for extinction and for the distances to their host galaxies
(distance to SN 2013df, above; 3.6 Mpc for SN 1993J, \citealt{freedman01}; 8.4 Mpc for SN 2011dh,
\citealt{vandyk13b}; and 9.6 Mpc for SN 2008ax, \citealt{pastorello08}).
We discuss below our estimate of the extinction for SN 2013df.
\citet{ergon13} have estimated the total extinction for SN 1993J and SN 2011dh 
to be $A_V \approx 0.53$ and $A_V \approx 0.22$ mag, respectively, and \citet{pastorello08} estimate
$A_V=0.93$ mag for SN 2008ax
(all assuming a \citealt{cardelli89} reddening law and $R_V=3.1$). 

SN 2013df appears to be the least luminous of all of these SNe. 
The differences in the peak luminosity likely arise from differences in the  
synthesized $^{56}$Ni mass \citep[e.g.,][]{perets10}.
For comparison, the $^{56}$Ni produced in the comparison SNe was $\sim 0.07$--$0.15\ M_{\odot}$,
$\sim 0.10$--$0.15\ M_{\odot}$, and $\sim 0.06\ M_{\odot}$ in 
SN 2008ax \citep{pastorello08,taubenberger11}, SN 1993J \citep{young95,richardson06}, and
SN 2011dh \citep{bersten12}, respectively.

From the overall comparison of the SN 2013df light curves in 
Figures~\ref{figlc} and \ref{figlcabs} with the light curves of SNe 1993J and 2011dh, we can make an 
estimate of the explosion date based, in particular, on when
SN 2013df appears to have reached secondary maximum in $BVRI$. 
We find that all of the light curves indicate that this date is JD 2,456,447.8 $\pm 0.5$, or about June 4.3
(indicated in Fig.~\ref{figlc}).
This is certainly consistent with the earliest discovery epoch of June 7.87 (JD 2,456,451.37). 
We are unable to constrain the explosion date based on KAIT SN search monitoring, since the 
observation of the host galaxy with KAIT
prior to discovery was on May 25 (i.e., about 10 days before discovery), at a limiting 
(unfiltered) magnitude of 18.5.

\subsection{Spectrum}

\begin{figure}
\figurenum{4}
\plotone{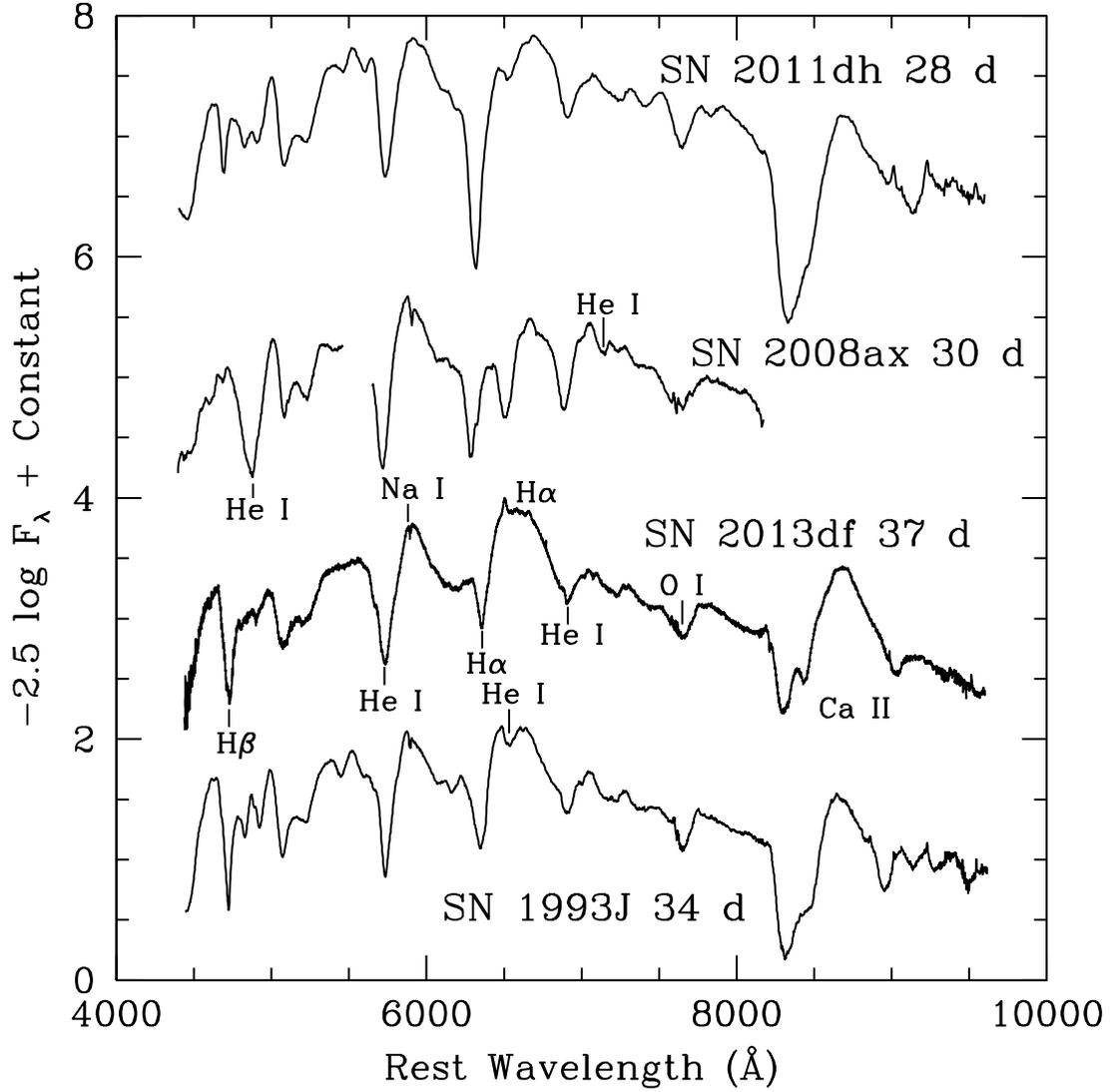}
\caption{Optical spectrum of SN 2013df obtained on 2013 July 11.2 with DEIMOS on 
the Keck II 10-m telescope. Also shown for comparison are the spectra, at approximately
the same age, of SN 1993J from 1993 April 30 \citep{filippenko93}, SN 2008ax 
from 2008 April 2 \citep{pastorello08}, and SN 2011dh from 2011 June 29 (unpublished; 
from WISeREP, \citealt{yaron12}). Several spectral features are indicated.
\label{figspec}}
\end{figure}

In Figure~\ref{figspec} we show the Keck/DEIMOS spectrum of SN 2013df and 
a comparison with spectra of the SNe~IIb 1993J \citep{filippenko93}, 2008ax 
\citep[][ obtained from SUSPECT\footnote{http://suspect.nhn.ou.edu/$\sim$suspect/.}; the data are also available at WISeREP\footnote{\citet{yaron12}, http://www.weizmann.ac.il/astrophysics/wiserep/.}]{pastorello08},
and 2011dh (available from WISeREP), all at nearly the same age.
The SN 1993J explosion date was 1993 March 27.5 \citep{lewis94}, 
so the spectrum of SN 1993J from April 30 shown in the figure is at age $\sim 34$ d.
The age of the spectrum of SN 2008ax is $\sim 30$ d \citep{pastorello08}.
The age of the SN 2011dh spectrum is $\sim 28$ d \citep[assuming an explosion date of
2011 May 31.275;][]{arcavi11}.
From our estimate of the explosion date, above, 
the SN 2013df spectrum shown in Figure~\ref{figspec} was obtained on day $\sim 37$. 
This spectrum bears a greater similarity with that of SN 1993J than SN 2008ax. 
Several of the He~{\sc i} lines, notably the $\lambda$6678 line, atop the broad H$\alpha$ emission
profile, appear to have been stronger in the 
SN 1993J spectrum (even at a somewhat earlier age) than for SN 2013df.
This could imply that the H layer for SN 2013df
may have been more substantial (larger mass, larger radius) at explosion, 
as suggested as well by the post-breakout light-curve behavior of SN~2013df, relative to SN 1993J.

\begin{figure}
\figurenum{5}
\plotone{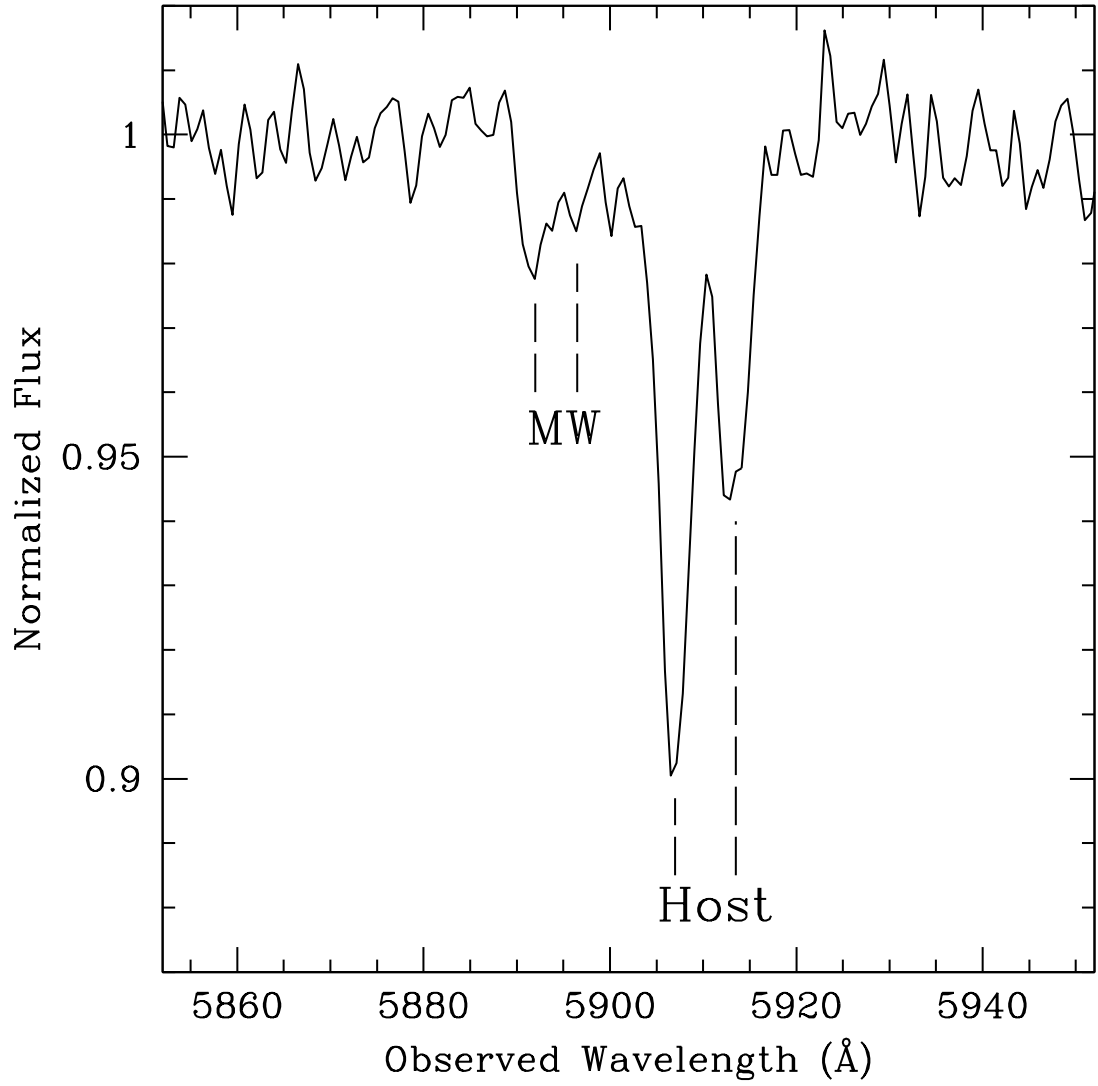}
\caption{Portion of the spectrum of SN 2013df shown in Figure~\ref{figspec}, after normalization of
the continuum, focusing on the region including the 
interstellar Na~{\sc i}~D $\lambda\lambda$5890.0, 5895.9 absorption features. 
The possible feature
associated with the Galactic component is indicated by the dashed lines as ``MW,'' and the feature 
(corresponding to a redshift of 0.002874) associated with the host galaxy is indicated as ``Host.''
\label{fignaid}}
\end{figure}

Visible in the spectrum of SN~2013df are absorption features due to Na~{\sc i}~D. 
In Figure~\ref{fignaid} we show these features after the overall continuum in the spectrum has been
normalized. One pair of lines appears to be weakly visible at effectively zero redshift,
which would correspond to the Galactic foreground extinction contribution; 
these are indicated as ``MW'' in the figure.
Another pair is far more distinct, at a redshift of 0.002874, 
indicated as ``Host'' in the figure. This is consistent with the redshift for NGC~4414,  0.002388, given 
by NED\footnote{The NASA/IPAC Extragalactic Database, http://nedwww.ipac.caltech.edu.}.
We have measured the equivalent width (EW) of each of the Na~{\sc i} features, $D_1$ and $D_2$,
for each system to be
EW($D_2$) = $0.097 \pm 0.011$ \AA\ and EW($D_1$) = $0.049 \pm 0.001$ \AA\ from the Galactic 
system, 
for a total EW(Na~{\sc i}) = 0.146 \AA. For the host-galaxy system,
EW($D_2$) = $0.433 \pm 0.002$ \AA\ and EW($D_1$) = $0.259 \pm 0.001$ \AA,
for a total EW(Na~{\sc i}) = 0.691 \AA.
The ratio EW($D_2$)/EW($D_1$) is 1.98 for the Galactic component and 1.67 for the host-galaxy
component. The ratio of the oscillator strengths of these two lines is 2.0; this intrinsic value is approached
only at the lowest optical depths \citep{munari97}.
We can see that the EW(Na~{\sc i}) [host] is $\sim$4.7 times that of EW(Na~{\sc i}) [Galactic].
If we assume the Galactic foreground extinction estimated by \citet[][ $A_V=0.053$ mag;]{schlafly11}, 
then from the ratio of these two components, it is plausible that the 
extinction internal to the host galaxy is $A_V=0.25$ mag. 

\citet{poznanski11} have stressed that one cannot accurately infer interstellar visual extinction from 
EW(Na~{\sc i}) in low-resolution spectra. The resolution of the Keck DEIMOS spectrum  may be just
at the limit, however, where we can use the more accurate relations from \citet{poznanski12}. If we 
apply
their Equation 7 for EW($D_2$), we obtain $E(B-V)=0.11 \pm 0.04$ mag; from their Equation 8 for
EW($D_1$), we obtain $E(B-V)=0.08 \pm 0.04$ mag; and from their Equation 9 for the total 
EW($D_1+D_2$), $E(B-V)=0.09 \pm 0.01$ mag. The uncertainty-weighted mean from all three 
estimates is then $E(B-V)=0.09 \pm 0.01$ mag. If we assume the \citet{cardelli89} reddening law,
with $R_V=3.1$, then the host-galaxy extinction is $A_V=0.28 \pm 0.04$ mag. (We note that
\citealt{poznanski12} assume the older, somewhat higher values of Galactic extinction from 
\citealt{schlegel98}.) This estimate of the  host-galaxy extinction is quite consistent with the one we 
made above, which we adopt. We therefore assume hereafter that the total (Galactic foreground plus 
host galaxy) extinction for SN 2013df is $A_V=0.30$ mag. We adopt an uncertainty in the extinction of 
$\pm 0.05$ mag.

\subsection{Progenitor}

We display in Figures~\ref{figprog}a and \ref{figprog}b the subregion 
of the WFPC2 mosaic around the SN position at 
F555W and F814W, respectively. We show the combination of all the WFC3 
exposures in Figure~\ref{figprog}c.
From a precise comparison between the pre-explosion WFPC2 images and post-explosion WFC3 
images, we can identify the SN progenitor. We had previously 
attempted an identification of the star \citep{vandyk13a}, but this was employing much 
lower-resolution, ground-based images of the SN. 
None of the three objects that we had previously nominated as candidates, numbered in 
Figure~\ref{figprog},
turned out to be the likely SN progenitor. Using 11 stars in common between the WFPC2 and WFC3 
images,
we astrometrically registered the images to an accuracy of 0.11 drizzled WFPC2 pixel, 
or 5.5 milliarcsec, and found, instead, that the SN 
corresponds to the position of the star indicated by the tick marks in Figures~\ref{figprog}a and \ref{figprog}b. 
We therefore identify this star as the likely progenitor of SN~2013df. 

\begin{figure}
\figurenum{7}
\plotone{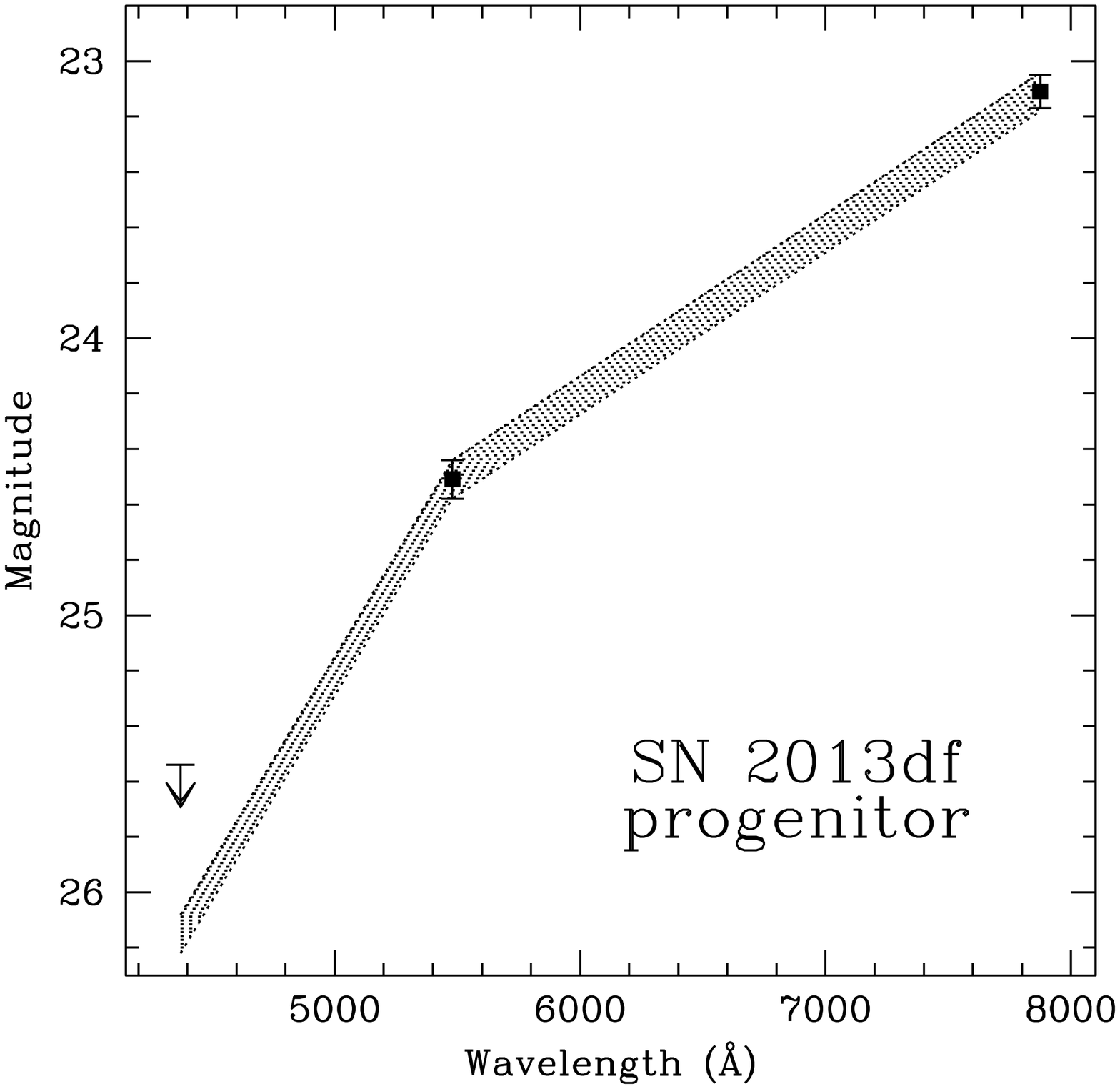}
\caption{Spectral energy distribution of the probable progenitor star of SN 2013df, based on
the photometry presented in Table~\ref{hstphot}.
Shown for comparison is synthetic photometry derived from a
MARCS \citep{gus08} supergiant model atmosphere at 
solar metallicity with effective temperature 4250 K, reddened assuming
$A_V=0.30$ mag and the \citet{cardelli89} reddening law. The model has
been normalized at $V$. The hashed region in the diagram represents the range in brightness for 
the model within the total uncertainty in the observed $V$ magnitude.\label{figsed}}
\end{figure}

We extracted photometry for the source from the pre-explosion WFPC2 images
using Dolphot v2.0 \citep{dolphin00}. (We disregarded the short F439W, F606W, and F814W 
exposures, since the signal-to-noise ratio is quite low in these.) The progenitor location is in the 
WFPC2 chip 3.
The Dolphot output indicates that the progenitor is likely stellar, with an ``object type'' of 1; the
``sharpness'' parameter, at $-0.37$, is slightly outside the acceptable range ($-0.3$ to +0.3)
for a ``good star'' \citep{dolphin00,leonard08}. However, if we run Dolphot on the F555W and F814W 
images only (excluding the F439W images), this parameter has a more acceptable value of $-0.25$. 
We therefore consider it most likely that this is a well-resolved stellar object.
The star is not detected by Dolphot in the F439W images (nor is it detectable upon visual
inspection of the images).  However, we estimate a $3\sigma$ upper limit to its detection.
The Dolphot output 
automatically includes the transformation from {\sl HST\/} flight-system magnitudes 
to the corresponding Johnson-Cousins magnitudes \citep{holtzman95}.
We present the  Dolphot results for the progenitor in Table~\ref{hstphot}. 
We also show the resulting spectral energy distribution (SED) of the star in Figure~\ref{figsed}.

To our knowledge, no measurement exists of the metallicity in the vicinity of the SN position (such as
from spectroscopy of nearby H~{\sc ii} regions). So, we are unable to quantify accurately 
the metallicity at the SN site with available data. 
However, we can provide at least an approximate estimate, based on
the assumption that the abundance gradient in the host spiral galaxy follows the same behavior as
for the sample of spirals analyzed by \citet{zaritsky94}. After deprojecting NGC 4414 from its
inclination and position angle \citep[56\arcdeg\ and 160\arcdeg, respectively;][]{jarrett00},
we find that SN 2013df occurred 
$\sim 31\arcsec$ (or $\sim 2.5$ kpc) from the nucleus. 
From the \citeauthor{zaritsky94}~nuclear abundance and gradient, then, 
the oxygen abundance (a proxy for metallicity) at the SN site is 
$12 + \log{[{\rm O/H}]} \approx 8.8$, which is comparable to the solar abundance
\citep[$8.69{\pm}0.05$;][]{asplund09}.
Therefore, it is reasonable to assume that the SN site is of roughly solar metallicity.

We compared the observed SED of the progenitor with synthetic SEDs
derived from MARCS \citep{gus08} supergiant model stellar 
atmospheres\footnote{http://marcs.astro.uu.se/.}. 
The model atmospheres are spherical, 
with standard composition at solar metallicity, surface gravity $\log g=1.0$, and 
microturbulence of 5 km s$^{-1}$. The spherical-geometry model atmospheres were computed, generally,
for stars with masses 0.5, 1.0, 2.0, 5.0, and 15 $M_{\odot}$; we chose models for this
last mass as most appropriate for the likely massive progenitor, given what has been inferred for
the progenitor initial masses for other SNe~IIb.
We found that the synthetic $V-I$ colors, in particular, derived from models at a given temperature are 
essentially mass-independent (they differ by $\sim 0.03$ mag).
The models were reddened based on the assumed total extinction for
SN 2013df, above, assuming the \citet{cardelli89} reddening law, and were normalized at $V$.
The model that provided the best comparison
with the photometry has $T_{\rm eff}=4250$ K, which we show in Figure~\ref{figsed}. 
Warmer and cooler models did not compare well with the observations.
We therefore adopt this effective temperature for the progenitor, with a generous uncertainty of 
$\pm 100$ K.

The absolute $V$ magnitude of the progenitor, corrected for the assumed 
distance modulus and extinction to the SN, is $M_V^0=-6.89 \pm 0.10$. 
The uncertainty arises from the uncertainties in the photometry, the extinction, and the distance modulus, 
all added in quadrature. The uncertainty
in the transformation from flight-system to Johnson-Cousins magnitudes is 
$\ll 0.01$ mag \citep{holtzman95}. 
The bolometric correction obtained from the MARCS model at $T_{\rm eff}=4250$ K is 
$BC_V=-0.71$ mag. 
The uncertainty in $BC_V$, resulting from the uncertainty in $T_{\rm eff}$, is 
$\sim 0.05$ mag.
Including this uncertainty, the bolometric magnitude is
$M_{\rm bol}=-7.60 \pm 0.15$. 
Assuming a bolometric magnitude
for the Sun of 4.74, this corresponds to a luminosity $L_{\rm bol}=10^{4.94 \pm 0.06}\ L_{\odot}$.
We show the locus for the progenitor in a Hertzsprung-Russell (HR) diagram in Figure~\ref{fighrd}.

\section{Discussion and Conclusions}

\begin{figure}
\figurenum{8}
\plotone{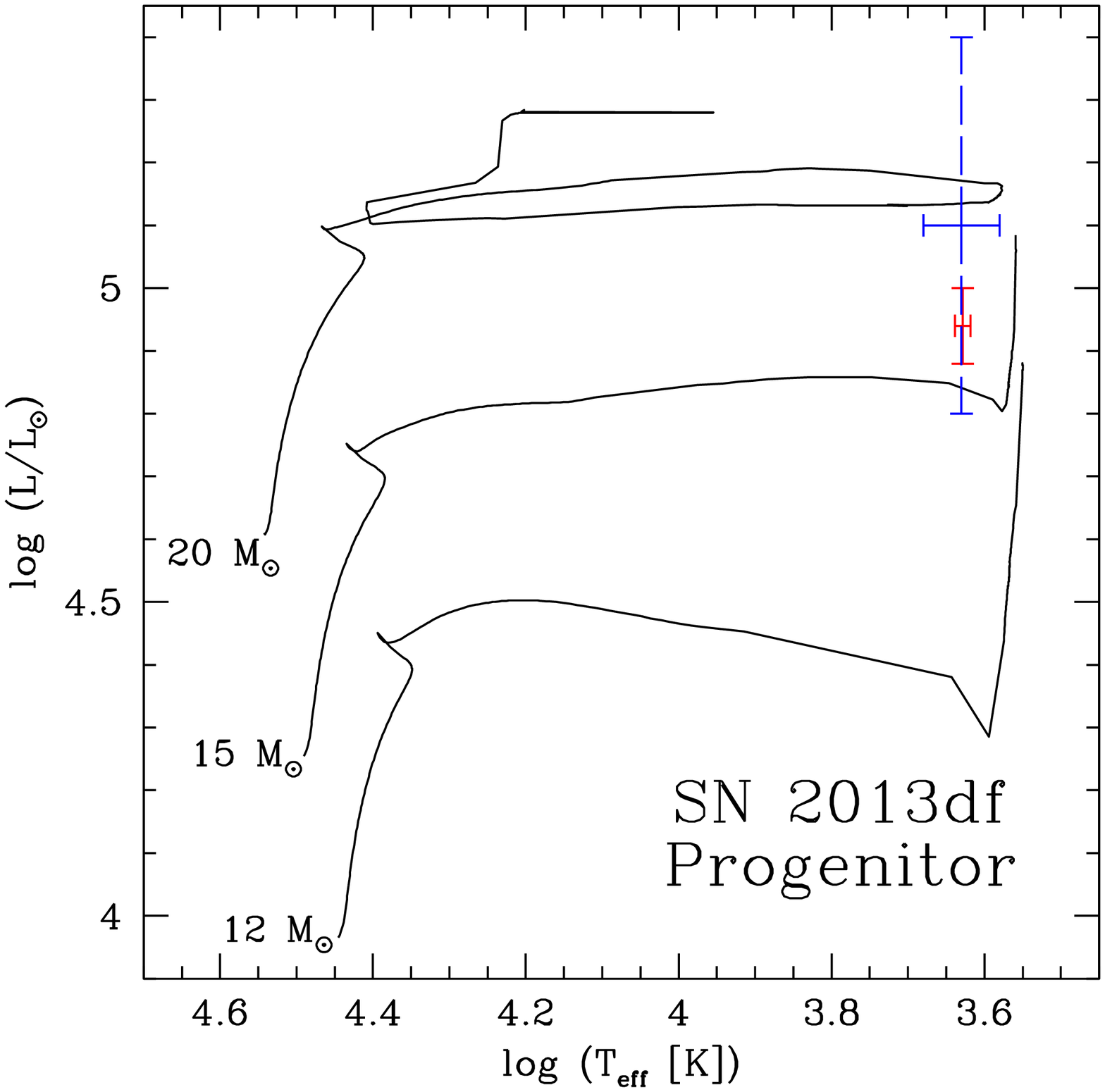}
\caption{Hertzsprung-Russell diagram, showing the locus of 
the probable progenitor of SN 2013df (solid symbol; colored red in the online version). 
For comparison we also illustrate the massive-star
evolutionary tracks with rotation from \citet{ekstrom12} at initial masses 
12, 15, and 20 M$_{\odot}$ (curves), as well as the locus of the SN 1993J progenitor
(dashed symbol; colored blue in the online version) from \citet{maund04}. \label{fighrd}}
\end{figure}

The SN 2013df progenitor's position in the HR diagram is significantly blueward of the RSG 
terminus of model massive-star evolutionary tracks with rotation, such as the 15 $M_{\odot}$ track
\citep{ekstrom12} in Figure~\ref{fighrd}. This indicates that the star likely has
 had its envelope somewhat stripped by some mechanism. 
The expectation so far for the progenitor scenario for SNe~IIb is an interacting massive
star binary system, such as for SN 1993J \citep[e.g.,][]{pod93,maund04} and SN 2011dh 
\citep{bersten12,benvenuto13}. (However, see \citealt{crockett08} for other possibilities regarding the
progenitor of SN 2008ax.)
A binary channel is also strongly favored for SNe~IIb, in general, based on statistical arguments accounting 
for their observed relative rate among core-collapse SNe \citep{smith11}, as well as on the small ejecta 
masses and H/He line ratios compared to detailed models \citep{dessart11,hachinger12}.
A blue binary companion is expected to survive the explosion for SNe~IIb \citep{maund04,benvenuto13}.
The locus of the SN 2013df progenitor in the HR diagram is quite similar to, albeit possibly
somewhat less luminous than (though within the uncertainties), that of the supergiant progenitor of 
SN 1993J \citep[][ see also Fig.~\ref{fighrd}]{maund04}.

We can attempt to assign an
initial mass to the progenitor, which, following \citet{vandyk11} for SN 2011dh, would have been
$\sim 16$--$17\ M_{\odot}$, based on comparing the locus in $T_{\rm eff}$ and $L_{\rm bol}$ to
the corresponding values of the 15 and 20 $M_{\odot}$ tracks in the figure.
On the other hand, adopting the approach of \citet{maund11}, of basing the likely initial mass on the
luminosity at the terminus of the available tracks, we estimate the mass to be 
$\sim 13$--$14\ M_{\odot}$. As \citeauthor{maund11}~(and references therein) pointed out,
the final luminosity at the top of the RSG branch is more relevant to the condition of the
yellow supergiant. The timescale for the transition redward for massive stars across the HR diagram 
is comparatively fast, and the stars will subsequently sit at the RSG locus for the majority of their 
post-main-sequence evolution.  In a binary model, mass transfer will truncate the outer radius
of the donor star, resulting in a hotter photosphere; however, the core nuclear-burning luminosity will be
relatively unaffected.  Note that, if the progenitor experienced mass transfer in a binary,
substantial mass stripping can lower the core temperature and, hence, lower the surface luminosity. 
So, the progenitor could have been somewhat more massive initially, perhaps $\sim 15\ M_{\odot}$.

The effective radius of the SN 2013df progenitor 14 yr before explosion, based on our estimates
of $T_{\rm eff}$ and $L_{\rm bol}$, was $R_{\rm eff} = 545 \pm 65\ R_{\odot}$.
Unfortunately, we cannot produce an independent estimate from the light curves
using the relations between SN progenitor radius and photospheric 
temperature by \citet{chevalier08}, \citet{nakar10}, and \citet{rabinak11},
since these relations tend to break down only a few days after 
explosion, and the earliest set of photometric data points for SN 2013df is at age $\sim$10 d.
This star was evidently more extended than the progenitor of SN~2011dh --- 
the yellow supergiant progenitor of SN 2011dh modeled by \citet{bersten12} has radius 
$R \approx 200\ R_{\odot}$.
The later post-shock breakout cooling for SN 2013df  implies
that its progenitor may have had a larger radius
than that of the SN 1993J progenitor. However, both \citet{maund04} and \citet{vandyk13b} 
calculated that the progenitor of SN 1993J had a radius of $\sim 600\ R_{\odot}$, which is 
somewhat greater than, although within the uncertainties of,  what we have estimated for the SN 
2013df progenitor.
In order for the SN 1993J progenitor to have had a smaller effective radius, 
it must have been hotter ($> 4265$ K) and of lower
luminosity ($< 10^{5.1}\ L_{\odot}$) than currently understood \citep[e.g.,][]{maund04}.
To have a larger radius, 
the SN 2013df progenitor would have had to be cooler and also potentially more luminous. 
However, to be cooler would require the extinction to be even lower than the small amount that we have
estimated. Additionally, the somewhat larger bolometric correction with the lower temperature would be 
offset by the smaller extinction correction, leading, ultimately, to very little change in the estimated 
luminosity.
At the least, we can say that the SN 2013df and SN 1993J progenitor stars appear to have been relatively 
comparable in nature.

In summary,
we have shown that SN~2013df in NGC 4414 is a SN~IIb, based on early-time photometry and
spectroscopy, and that its properties are most similar to those of SN~1993J. However, the mass of 
$^{56}$Ni synthesized in the explosion
likely differed from that in SNe~1993J, 2008ax, and 2011dh. Furthermore, 
the light curves of
SN~2013df --- specifically, the late cooling from post-shock-breakout maximum --- indicate that its
progenitor must have been more extended in radius than that of SN 1993J. 
Using archival pre-SN, high spatial resolution {\sl HST\/} images, together with
{\sl HST\/} images of SN~2013df, we have identified a star that
we consider to be the likely SN progenitor; it had
$T_{\rm eff}\approx 4250$ K and $L_{\rm bol}\approx 10^{4.94}\ L_{\odot}$. 
Ultimately, verification of this identification of the SN~2013df progenitor should occur when the SN
itself has greatly faded, as has been done for the SNe~IIb 1993J \citep{maund09} 
and 2011dh \citep{ergon13,vandyk13b}. This can be accomplished with high-quality images obtained 
with {\sl HST}. Deep, very late-time {\sl HST\/} imaging in the blue or
ultraviolet could reveal a putative blue binary companion star at the SN location, although the fact 
that SN 2013df is at least a factor of two more distant than both SNe 2011dh and 1993J could make 
such a detection a more challenging prospect. 

\acknowledgments 

We thank the referee for their comments.
This work  
is based in part on observations made with the NASA/ESA
{\it Hubble Space Telescope}, obtained from the Data Archive at the
Space Telescope Science Institute (STScI), which 
is operated by the Association of Universities for Research in Astronomy (AURA), Inc.,
under NASA contract NAS5-26555. Some of the data presented herein were obtained at the W. M. 
Keck Observatory, which is operated as a scientific partnership among the California Institute of 
Technology, the University of California and NASA; the
observatory was made possible by the generous financial support of the W. M. Keck Foundation.
KAIT and its
ongoing research were made possible by donations from Sun
Microsystems, Inc., the Hewlett-Packard Company, AutoScope
Corporation, Lick Observatory, the NSF, the University of California,
the Sylvia \& Jim Katzman Foundation, and the TABASGO Foundation. 
We thank the RATIR instrument
team and the staff of the Observatorio Astron\'omico Nacional
on Sierra San Pedro M\'artir. RATIR is a collaboration between
the University of California, the Universidad Nacional
Auton\'oma de M\'exico, NASA Goddard Space Flight Center,
and Arizona State University, benefiting from the loan of
an H2RG detector from Teledyne Scientific and Imaging.
RATIR, the automation of the Harold L.~Johnson Telescope
of the Observatorio Astron\'omico Nacional on Sierra San Pedro
M\'artir, and the operation of both are funded by the partner
institutions and through NASA grants NNX09AH71G,
NNX09AT02G, NNX10AI27G, and NNX12AE66G, CONACyT
grants INFR-2009-01-122785, UNAM PAPIIT grant
IN113810, and a UC MEXUS-CONACyT grant.
Support for
this research was provided by NASA through grants GO-12888 and GO-13030 
from STScI. A.V.F. and
his group at UC Berkeley also wish to acknowledge generous support from
Gary and Cynthia Bengier, the Richard and Rhoda Goldman Fund, the
Christopher R. Redlich Fund, the TABASGO Foundation, and NSF 
grant AST-1211916.  
Research by A.G. is supported by the EU/FP7 via ERC grant n~307260, ``The Quantum Universe'' I-Core 
program by the Israeli Committee for planning and budgeting, the ISF, GIF, and Minerva grants, and the 
Kimmel award. S.B. is supported by the Ilan Ramon Fellowship from ISA.

{\it Facilities:} \facility{HST(WFPC2)}, \facility{HST(WFC3)}, \facility{Keck}, \facility{KAIT}, \facility{RATIR}.


\begin{deluxetable}{cccccc}
\tablewidth{4.6truein}
\tablecolumns{6}
\tablecaption{{\sl HST\/} Photometry of the SN 2013df Progenitor\tablenotemark{a}\label{hstphot}}
\tablehead{
\colhead{F439W\tablenotemark{b}} & \colhead{$B$\tablenotemark{b}} & 
\colhead{F555W} & \colhead{$V$} & 
\colhead{F814W} & \colhead{$I$} \\
\colhead{(mag)} & \colhead{(mag)} & \colhead{(mag)}
& \colhead{(mag)} & \colhead{(mag)}
& \colhead{(mag)}}
\startdata
$>$25.65 & $>$25.54 & 24.535(071) & 24.515 & 23.144(055) & 23.106 \\
\enddata
\tablenotetext{a}{Uncertainties $(1\sigma)$
are given in parentheses as millimagnitudes.}
\tablenotetext{b}{$3\sigma$ upper limit.}
\end{deluxetable}

\clearpage

\clearpage

\clearpage

\clearpage

\clearpage

\clearpage

\begin{figure}
\figurenum{6}
\plotone{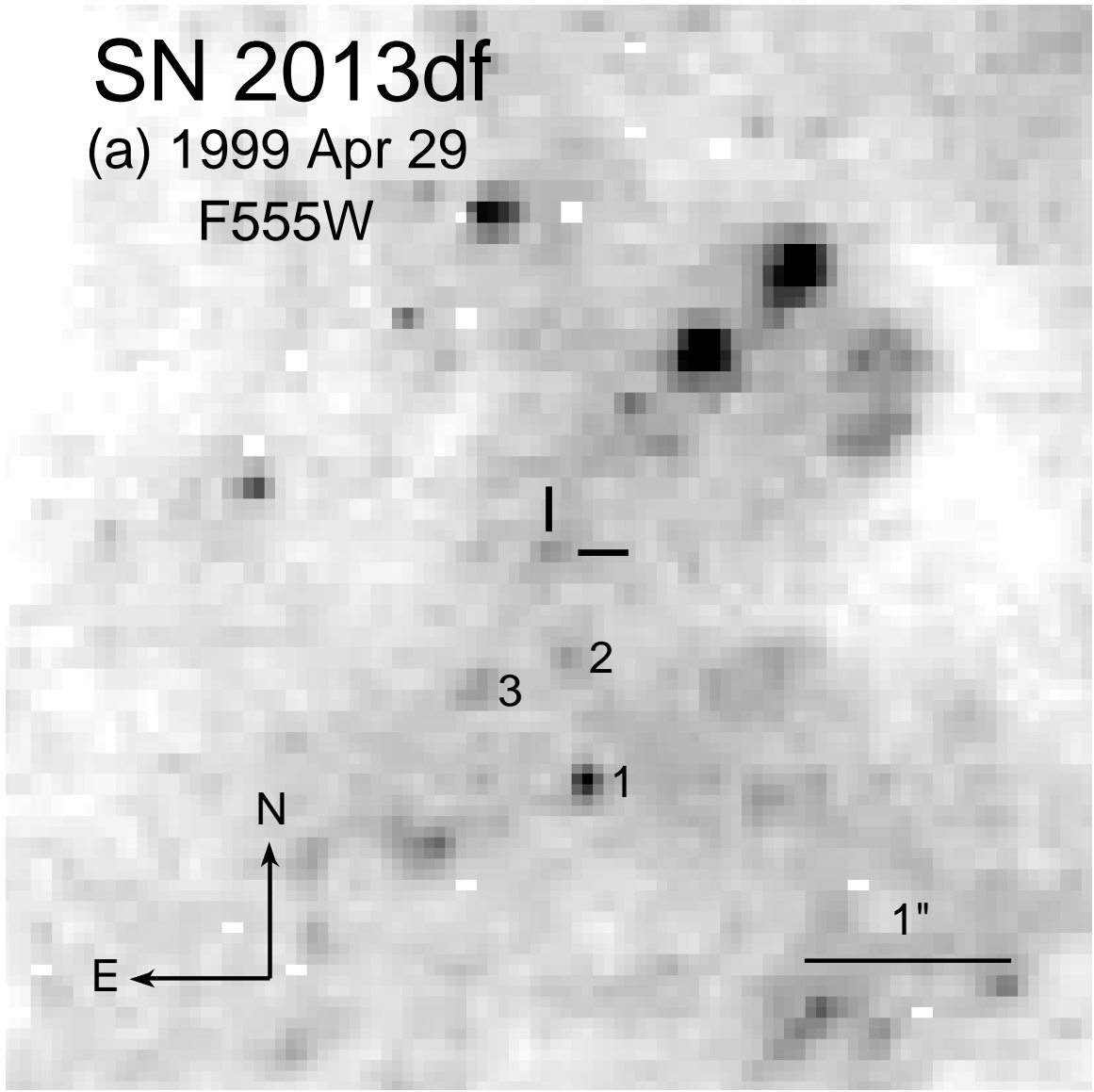}
\caption{A portion of the archival {\sl HST} WFPC2  images of NGC 4414 from 1999 in 
(a) F555W and (b) F814W. The likely progenitor of SN 2013df is indicated by tick marks. 
(c) A portion of the {\sl HST} WFC3 F555W image of SN 2013df, to the same scale and
orientation. The SN is indicated by tick marks. We have astrometrically registered the
WFPC2 and WFC3 images to an accuracy of 5.5 milliarcsec. The three progenitor candidates
initially identified by \citet{vandyk13a} are indicated (``1,'' ``2,'' ``3'') in panel (a).
North is up, east is to the left.\label{figprog}}
\end{figure}

\clearpage

\begin{figure}
\figurenum{6}
\plotone{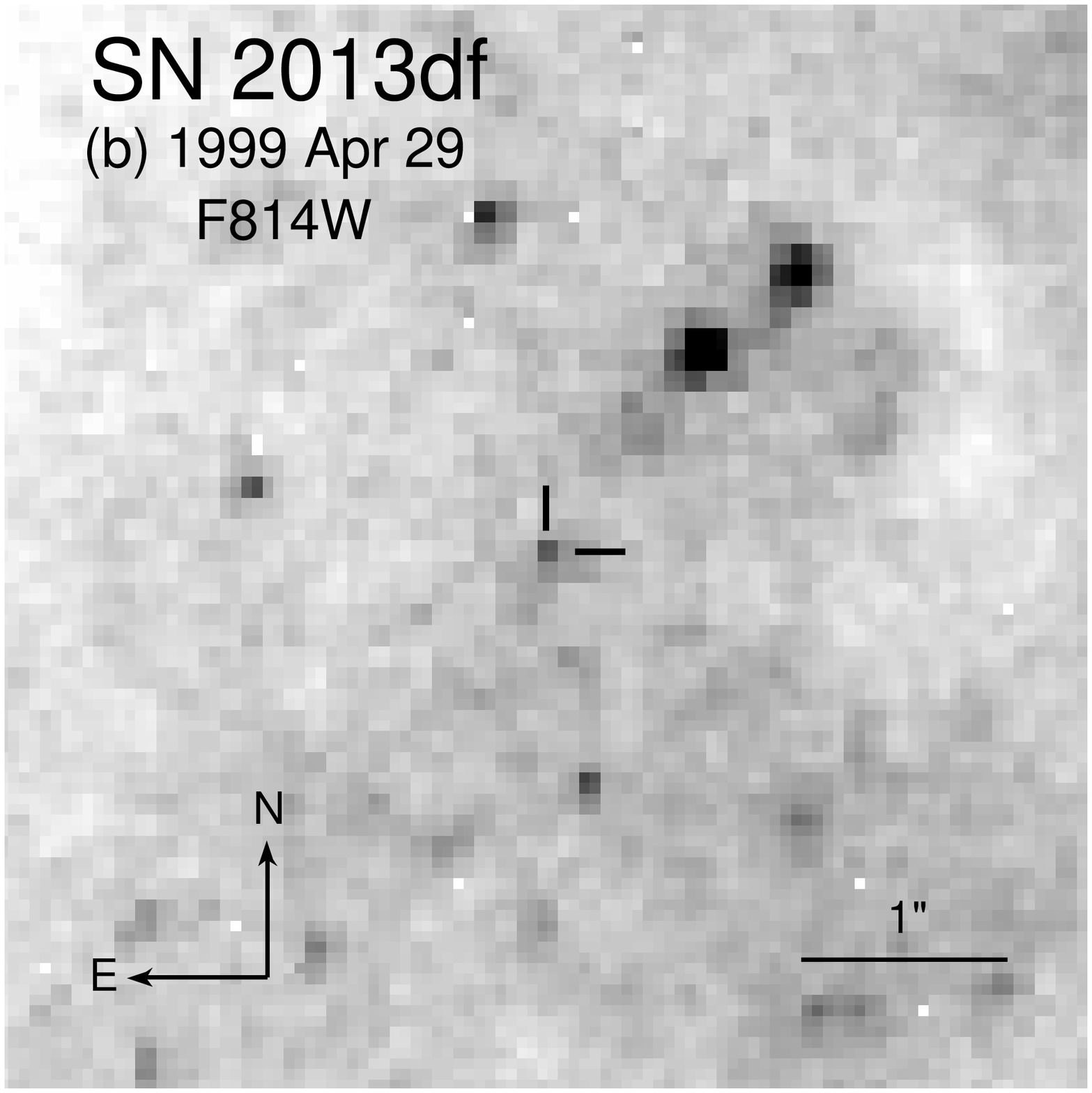}
\caption{(Continued.)}
\end{figure}

\clearpage

\begin{figure}
\figurenum{6}
\plotone{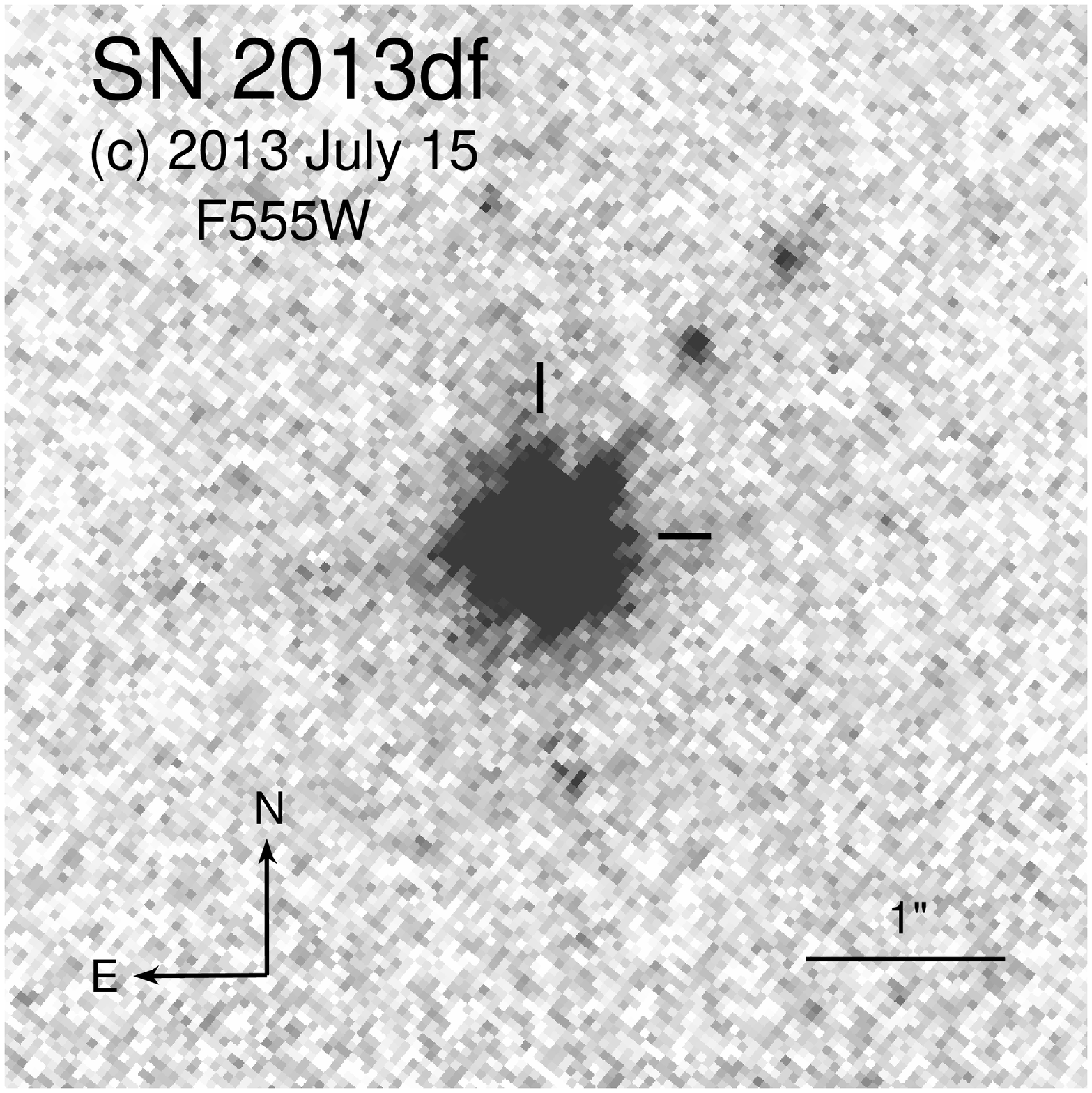}
\caption{(Continued.)}
\end{figure}

\clearpage

\clearpage

\end{document}